\documentclass[journal=jacsat,manuscript=article,layout=onecolumn]{achemso}

\usepackage{graphicx}
\usepackage{dcolumn}
\usepackage{bm}
\usepackage{verbatim}
\usepackage[utf8]{inputenc}
\usepackage[T1]{fontenc}
\usepackage{mathptmx}
\usepackage{mathtools}
\usepackage{etoolbox}
\usepackage{caption}
\usepackage{subcaption}

\newcommand{\SiO}{\mathrm{SiO}_2}

\newcommand{\Tsample}{T_{\mathrm{sample}}}
\newcommand{\dTdev}{\Delta T_{\mathrm{sample}}} 
\newcommand{\dTdc}{\Delta T_{\mathrm{DC}}} 
\newcommand{\dTacOne}{\Delta T_{\mathrm{AC,1\omega}}} 
\newcommand{\dTacTwo}{\Delta T_{\mathrm{AC,2\omega}}} 
\newcommand{\dTacnw}{\Delta T_{\mathrm{AC,}n\omega}} 
\newcommand{\dTdcPeltier}{\Delta T_{\mathrm{DC,Peltier}}} 
\newcommand{\dTacPeltier}{\Delta T_{\mathrm{AC,Peltier}}} 
\newcommand{\betaN}{\beta_n} 
\newcommand{\Vdevtot}{V_{\mathrm{dev,tot}}} 
\newcommand{\Vdevdc}{V_{\mathrm{dev,DC}}} 
\newcommand{\Vdevac}{V_{\mathrm{dev,AC}}} 
\newcommand{\Rdev}{R_{\mathrm{dev}}}
\newcommand{\Pdev}{P_{\mathrm{dev}}}
\newcommand{\Pdevdc}{P_{\mathrm{dev,DC}}}
\newcommand{\PdevacOne}{P_{\mathrm{dev,AC1\omega}}}
\newcommand{\PdevacTwo}{P_{\mathrm{dev,AC2\omega}}}

\newcommand{\Vtot}{V_{\mathrm{tot}}^{\mathrm{WS}}}
\newcommand{\Rsensor}{R_{\mathrm{sen}}}
\newcommand{\Rseries}{R_{\mathrm{series}}}
\newcommand{\IZero}{I_{\mathrm{0}}}
\newcommand{\Vsensor}{V_{\mathrm{sensor}}} 
\newcommand{\VZero}{V_{\mathrm{0}}} 
\newcommand{\dV}{\Delta V} 
\newcommand{\dVdc}{\Delta V_{\mathrm{DC}}} 
\newcommand{\dVacOne}{\Delta V_{\mathrm{AC,1\omega}}} 
\newcommand{\dVacTwo}{\Delta V_{\mathrm{AC,2\omega}}} 
\newcommand{\dVacnw}{\Delta V_{\mathrm{AC,}n\omega}} 
\newcommand{\dVacOneC}{\Delta V_{\mathrm{AC,1\omega,C}}} 
\newcommand{\RZero}{R_{\mathrm{0}}}
\newcommand{\dRdc}{\Delta R_{\mathrm{DC}}} 
\newcommand{\dRac}{\Delta R_{\mathrm{AC}}} 
\newcommand{\dRacOne}{\Delta R_{\mathrm{AC,1\omega}}} 
\newcommand{\dRacTwo}{\Delta R_{\mathrm{AC,2\omega}}} 
\newcommand{\Rcl}{R_{\mathrm{cl}}}
\newcommand{\Rts}{R_{\mathrm{ts}}}
\newcommand{\Tsensor}{T_{\mathrm{sen}}} 
\newcommand{\dTsensor}{\Delta T_{\mathrm{sen}}} 
\newcommand{\dTsensordc}{\Delta T_{\mathrm{sen,DC}}} 
\newcommand{\dTsensoracOne}{\Delta T_{\mathrm{sen,AC,1\omega}}} 
\newcommand{\dTsensoracTwo}{\Delta T_{\mathrm{sen,AC,2\omega}}} 
\newcommand{\TsensorZero}{T_{\mathrm{sen,0}}}
\newcommand{\dTsensorZero}{\Delta T_{\mathrm{sen,0}}}
\newcommand{\Pel}{P_{\mathrm{el}}}
\newcommand{\PelZero}{P_{\mathrm{el,0}}}
\newcommand{\Peldc}{P_{\mathrm{el,DC}}} 
\newcommand{\dPelac}{\Delta P_{\mathrm{el,AC}}} 
\newcommand{\dPelacOne}{\Delta P_{\mathrm{el,AC, 1\omega}}} 
\newcommand{\dPelacTwo}{\Delta P_{\mathrm{el,AC, 2\omega}}} 
\newcommand{\Qts}{\dot{Q}_{\mathrm{ts}}} 
\newcommand{\Qtsdc}{\dot{Q}_{\mathrm{ts,DC}}} 
\newcommand{\Qcl}{\dot{Q}_{\mathrm{cl}}} 
\newcommand{\QclZero}{\dot{Q}_{\mathrm{cl,0}}} 
\newcommand{\Qcldc}{\dot{Q}_{\mathrm{cl,DC}}} 
\newcommand{\Plaser}{P_{\mathrm{laser}}} 

\newcommand{\dTdccal}{\Delta T_{\mathrm{DC}}^{\mathrm{cal}}} 
\newcommand{\dTdcsthm}{\Delta T_{\mathrm{DC}}^{\mathrm{SThM}}} 
\newcommand{\micrometer}{\mathrm{\mu m}}
\newcommand{\nanometer}{\mathrm{nm}}
\newcommand{\RdevZero}{R_{\mathrm{dev,0}}}

\newcommand{\An}{A_n}

\newcommand{\Ainf}{A_{\mathrm{\infty}}}
\newcommand{\Bn}{B_n}

\newcommand{\onehalf}{\frac{1}{2}}

\newcommand{\TRT}{T_{\mathrm{RT}}}

\newcommand{\Rboundary}{R_{\mathrm{boundary}}}
\newcommand{\rf}{r_{\mathrm{f}}}
\newcommand{\Tf}{T_{\mathrm{f}}}
\newcommand{\hk}{h_{\mathrm{k}}}
\newcommand{\RspreadTiN}{R_{\mathrm{spread,TiN}}}
\newcommand{\RspreadHfOTwo}{R_{\mathrm{spread,HfO2}}}
\newcommand{\rspreadHfOTwo}{r_{\mathrm{spread,HfO2}}}
\newcommand{\kappaTiN}{\kappa_{\mathrm{TiN}}}
\newcommand{\kappaHfOTwo}{\kappa_{\mathrm{HfO2}}}
\newcommand{\tHfOTwo}{t_{\mathrm{HfO2}}}
\newcommand{\tTiN}{t_{\mathrm{TiN}}}
\newcommand{\Rtot}{R_{\mathrm{tot}}}
   
\author{Nele Harnack}
\affiliation[]
{IBM Research Europe - Zurich, 8803 R\"uschlikon, Switzerland}
\author{Sophie Rodehutskors}
\affiliation[]
{IBM Research Europe - Zurich, 8803 R\"uschlikon, Switzerland}
\altaffiliation{Current address: Max Planck Institute for the Structure and Dynamics of Matter, 22761 Hamburg, Germany}
\author{Bernd Gotsmann}
\affiliation[]
{IBM Research Europe - Zurich, 8803 R\"uschlikon, Switzerland}
\email{bgo@zurich.ibm.com}

\title[]{Scanning Thermal Microscopy method for self-heating in non-linear devices and application to current filaments in resistive RAM}

\begin{document}

26th September 2024

\begin{abstract}
Resistive RAM (RRAM) devices are candidates for neuromorphic computing devices in which the functionality lies in the formation and reversible rupture and gap-closing of conducting filaments in insulating layers. To explore the thermal properties of these nanoscale filaments, Scanning Thermal Microscopy (SThM) can be employed. However, since RRAM devices, as well as many other neuromorphic device types, have a non-linear resistance-voltage relationship, the high resolution and quantitative method of AC-modulated SThM cannot readily be used. To this end, an extended non-equilibrium scheme for temperature measurement using SThM is proposed, with which the self-heating of non-linear devices is studied without the need for calibrating the tip-sample contact for a specific material combination, geometry or roughness. Both a DC and an AC voltage are applied to the device, triggering a periodic temperature rise, which enables the simultaneous calculation of the tip-sample thermal resistance and the device temperature rise. The method is applied to $\mathrm{HfO_2}$-based RRAM devices to extract properties like the number of current filaments, thermal confinement and thermal cross-talk. This approach could be applied to other thermometry techniques, including infrafred imaging and Raman thermometry. \newline \newline
orcid IDs: Nele Harnack https://orcid.org/0000-0002-7807-3950\\
keywords:
Scanning Thermal Microscopy, nanoscale thermometry, calibrationless, self-heating, quantitative, non-equilibrium, Resistive RAM
\end{abstract}

\section{\label{sec:level1}Introduction} 
The self-heating of integrated microelectronic devices during their operation has received considerable attention. On the one hand, the high current densities and subsequent temperature increase in devices and leads are known to be the main cause for their failure \cite{Pop2010}. 
On the other hand, the temperature rise can also be exploited to create device functionality.  
In either case, it is of high importance to understand and control self-heating in scaled devices, but several factors make it difficult to both predict and measure device properties: the multi-scale nature of heat transport, different self-heating effects, and different materials and material interfaces, which are often poorly characterised in their nanoscale thermal properties.

An important example are RRAM devices for neuromorphic computing architectures, which often use self-heating to change and store the resistive state \cite{Dittmann2021}. Given that the nonlinear thermal response is essential for device operation, it is surprising how few studies exist that access the temperature of such devices \cite{Deshmukh2022,Swoboda2023,West2023}, and how little is still known about their thermal properties, making it crucial to create high resolution and quantitative analysis methods.

Many methods of device thermometry with imaging capability have been applied to characterise self-heated devices, such as a range of scanning probe-based \cite{Borca-tasciuc2013} and optical methods like Raman thermography\cite{Kuball2016}. Scanning Thermal Microscopy\cite{Zhang,Gomes2015}
is particularly versatile as it can be applied to a large variety of samples under realistic operating conditions and routinely delivers high resolution \cite{Menges2012}. However, as we will discuss below, one issue of SThM is the difficulty of extracting quantitative information from the recorded electrical signals\cite{Borca-tasciuc2013} caused by the complex thermal coupling through the tip-sample contact \cite{Menges2016}. 
Methods to mitigate this issue include the null-point method \cite{Chung2010} or careful simulation of the tip-surface thermal contact \cite{Zhang, Menges2016}. Another versatile SThM method modulates the voltage (or current) applied to a device at a given frequency and uses phase sensitive detection of the thermal sensor at its harmonics to extract quantitative information, which has been used in SThM and thermoreflectance imaging \cite{Favaloro2015}. However, while applicability to linear devices has been shown \cite{Harzheim2018,Menges2016}, such AC modulation methods cannot be applied directly to non-linear electrical devices. 

In this paper, we demonstrate high-resolution imaging of hotspots to characterise current filaments in RRAM devices based on $\mathrm{HfO_2}$ layers sandwiched between TiN metal electrodes. This is enabled by an extension of the SThM methodology towards non-linear devices, obtaining a quantitative temperature in a single measurement without additional calibration. The method is sensitive enough to image self-heating effects through the cooling electrodes and enables extraction of important thermal dissipation characteristics. 

The paper is organised as follows. We first derive the equations for the new method,  discussing assumptions, strengths and limitations. To demonstrate that the method is quantitative, we use metal thin film resistors, of which the self-heating temperature can be determined independently. We then move on to motivate the study of $\mathrm{HfO_2}$-based RRAM devices and show the extraction of important properties through the  application of SThM. We continue with more common questions on the temperature and size of the current filament during operation before turning to time-dependent properties, which can now be accessed using the demonstrated method.

\section{An SThM technique for linear and non-linear devices}
\subsection{Equilibrium thermometry and non-equilibrium thermometry using SThM} 

\begin{figure*}[t]
    \centering
\includegraphics[width=\textwidth]{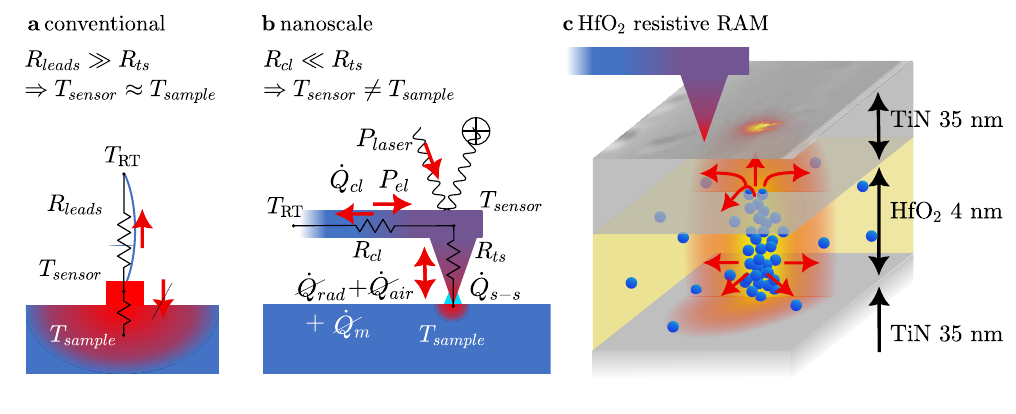}
\caption{(a) In conventional thermometry at large scales, the sensor is in thermal equilibrium with the sample to be measured, resulting from the thermal resistance of the sensor leads being small in comparison to the thermal resistance between sensor and sample. As a result, there is no heat flow between sample and sensor and the temperature of the sample can be read directly from the sensor temperature. 
\newline (b) In contact-based high-resolution techniques, the thermal resistance of the cantilever legs $\Rcl$ is lower than the resistance between tip and sample $\Rts$ in order to minimise the influence of the measurement on the sample temperature. To quantify a sample temperature correctly in non-equilibrium SThM, different sources of heat flux $\dot{Q}$ need to be considered. Vacuum-based measurements enable avoiding contributions from air conduction and minimise water meniscus.
\newline (c) $\mathrm{HfO_2}$-based RRAM device with conductive filament consisting of oxygen vacancies (indicated by blue spheres). Heat is dissipated from the filament in radial and axial direction, of which the resulting temperature at the top electrode is observable.}
\label{fig:first_illustration_thermal_resistances}
\end{figure*}

In Figure \ref{fig:first_illustration_thermal_resistances} a and b, conventional contact-based thermometry is compared to nanoscale temperature measurements. Depicted is the main difficulty in SThM: high resolution techniques do not operate in equilibrium of sensor and sample. The heat flux between sample and sensor is controlled by the thermal resistances between the sensor and the sample, $\Rts$. While the tip-cantilever thermal resistance remains constant, the tip-sample thermal resistance at the contact between probe and sample can be highly variable during the scan and cannot be assumed to be constant \cite{Menges2012}. It is affected by variations of contact area, given by device topography, surface roughness, the material combination's elasticity, the thermal conductivity of the sample and the thermal boundary resistance at the tip-sample interface. While features which have a similar size or are larger than the characteristic size of the tip-sample contact can be analysed using a calibration approach or using simulations, smaller features have to be addressed differently \cite{Swoboda2023}.

One proposed method is the use of a dual-scan mode \cite{Menges2016}, i.e. the combination of active and passive mode SThM to extract both tip-sample thermal resistance and temperature in a single measurement instead of two \cite{Kim2008a}. This is conceptually similar to varying the probe temperature to extrapolate to the null-point \cite{Chung2010}, as both methods are able to account for the variation in tip-sample thermal resistance. It has been successful in removing artifacts from measurements of various devices while maintaining a very high resolution\cite{Harzheim2018, Menges2016}. However, the methods proposed so far are only suitable for linear resistors in which the self-heating can easily be interpreted. For example, in a linear resistor modulated with a sinusoidal current or voltage at frequency $f$, the self-heating power due to Joule heating will be modulated at $2\,f$, while Peltier effects can be observed at $1\,f$ \cite{Gaechter2020}. If, in addition, the thermal resistance (coupling the device to its environment) is linear and the modulation small and sufficiently slow, the local temperature rise $\Delta T(x,y)$ will also be modulated with $2\,f$. This relation ($\Delta T \propto 2\,f$) translates readily to resistor networks and can account for heat spreading around the heat source in a device. When the resistor is non-linear, however, i.e. the current is no longer proportional to voltage, the simplified assumption cannot be made. Additionally, the investigation of devices at different offset voltages (also often referred to as 'DC bias') can reveal their characteristics at realistic operating conditions, opening up the possibility of characterising volatile states.

Here, we explore ways to use modulation techniques and phase-sensitive detection at different applied offset voltages, with the goal of extracting quantitative information from the temperature fields. To this end, a sensing and data analysis scheme for devices at constant offset voltage was developed based on the dual-scan method \cite{Menges2016}. The device is biased by a voltage with a constant offset (DC voltage) and an oscillating voltage (AC voltage). This induces an oscillating temperature in the device, which causes an oscillating heat flow between sample and probe and thus an oscillating resistance of the probe's integrated thermistor. As in the original dual-scan mode, the tip-sample thermal resistance is not assumed to be constant, but will be calculated independently for each image pixel.
The advantage of this modified dual-scan SThM method in comparison to the null-point method or modulating the heating current of the tip does not lie in simplifying assumptions or increasing elegance. It is, however, very easy to practically implement and in our tests shows better sensitivity and fewer artifacts. In the small signal limit, all methods coincide in their basic assumptions.

\subsection{Derivation of governing equations} \label{case_linear_const_current}

In this scheme, a device is electrically driven through modulation of a bias voltage (or, equivalently, a current) at modulation frequency $f$, including both AC and DC components:
\begin{align}
     \Vdevtot &= \Vdevdc  +\Vdevac \cdot \cos(2 \pi f t) \text{ with } \omega = 2 \pi f. \label{V_dev_offset}
\end{align}
The resulting Joule heating induces a temperature rise of the sample, which is also periodic:
\begin{align}     
\dTdev &= \dTdc + \dTacOne \cdot \cos(\omega t) \nonumber \\
&\quad+ \dTacTwo \cdot \cos(2 \omega t) \nonumber \\ 
&=\dTdc \left(1 + \beta_1 \cdot \cos(\omega t) +\beta_2 \cdot \cos(2 \omega t) \right) \label{eq:delta_T_sample}
\end{align}
\noindent{plus higher order terms. Here, $\beta_1$ and $\beta_2$ are introduced, and, as will be discussed below, can be determined independently. This leaves us with one unknown for the device, $\dTdc$, that is to be determined using this method for each pixel.} 

To analyse the relation between device temperature rise and operating conditions, one must choose the modulation frequency $\omega$ to be sufficiently low. In that case, the device maintains a steady state, i.e. a constant temperature within a time span that the applied voltage is almost constant. In other words, the modulation frequency should stay well below the device's cut-off frequency. Experimentally, this frequency can be determined by measuring the device's frequency response in the third harmonic \cite{Lu2001}. It follows that the device temperature rise is proportional to the power dissipated in the device, which allows us to compute the $\beta$ values. Additionally, the temperature rise in the device is assumed to be small, such that the thermal conductance of the device material and the thermal resistance of the tip-surface contact can be assumed independent of temperature and constant within one pixel. 

Next, the sensor's temperature out of contact and in contact with the sample are described. Out of contact with the sample, the sensor has a temperature that is elevated by $\dTsensorZero$ above room temperature $\TRT$, caused by self-heating through the voltage $\VZero$ applied to it via a Wheatstone-bridge. When the probe is brought into contact with the sample, its temperature changes due to the interaction with the sample. This induced temperature change is described by $\dTsensor$, which can have DC and/or AC contributions, depending on the temperature changes that occur in the sample. The total temperature in the cantilever is then 
\begin{align}
    \Tsensor &= \TRT + \dTsensorZero + \dTsensor \nonumber \\
    &= \TRT + \dTsensorZero + \dTsensordc
    \nonumber \\&\quad  + \dTsensoracOne \cdot \cos(\omega t)
    \nonumber \\&\quad + \dTsensoracTwo \cdot \cos(2 \omega t).  \label{eq:TsensorParts}
    \intertext{The sensor temperature rise is calculated via its calibrated temperature-dependent resistance and for small temperature changes can be linearised around the working point $\RZero = R(\dTsensorZero)$ to}
    &R = \RZero +\alpha \cdot \dTsensor. \label{eq:Rcantilever}
    \intertext{The resistance change is detected through the voltage drop over the cantilever}
    \Vsensor&=V_0 + \dVdc + \dVacOne \cdot \cos(\omega t) \nonumber \\
    &\quad + \dVacTwo \cdot \cos(2 \omega t). \label{eq:vsensor}
    \intertext{In a first approximation, the current can be treated as constant, because the periodic resistance changes $\dRac$ are small in comparison to the constant resistance of the heater $\dRdc$ and its series resistance $\Rseries$:}
    &I \simeq \Vtot / \left( \Rseries + \RZero +\dRdc \right)  \, ,\label{eq:const_current}
\end{align}
where $\Vtot$ is the voltage applied to the Wheatstone bridge. Then the temperature changes can be calculated directly from the corresponding constant or periodic voltage change $\dRdc = \dVdc / (I \cdot \alpha)$ . This assumption is written for simplicity and ease of understanding. For a more detailed treatment without assuming a constant current, which in many cases should be chosen over the simple approach, see the appendix.

To relate the local temperature change in the sample to the measured change in cantilever voltage, the continuity equation (energy conservation) of the cantilever is used, as depicted in figure \ref{fig:first_illustration_thermal_resistances}. 
\footnote{We neglect radiation or air conduction in our vacuum setup. If present, these can be taken into account in a single $\Qcl$ and eliminated in the measurement through the same calibration step used now}

First, for the tip out of contact with the sample, all the power dissipated in the sensor is conducted as heat along the cantilever legs (see figure \ref{fig:first_illustration_thermal_resistances}):
\begin{align}
    &\PelZero = \QclZero = \dTsensorZero / \Rcl. \label{eq:Qcl0}\\ 
    \intertext{Then, for the in-contact case}
    &\Pel =\Qcl + \Qts,\label{eq:ICenergy}
    \intertext{and}
    &\Qcl = (\dTsensorZero + \dTsensor)/\Rcl. \label{eq:Qcl}
    \intertext{Now the tip-sample contact provides an additional conduction channel, and the heat flux $\Qts$ can be described as} 
    &\Qts = (\TRT +\dTsensorZero + \dTsensor - \Tsample)/\Rts. \label{eq:Qts}
 \end{align}   
    
The electrical power of the heater can be calculated as $\PelZero = \VZero \cdot I$ in the out-of-contact case and $\Pel = (\VZero + \dV) \cdot I$ in the in-contact case.\newline
Plugging the terms for the device \eqref{eq:delta_T_sample} and the sensor temperature \eqref{eq:TsensorParts}, the sensor resistance and voltage \eqref{eq:Rcantilever}, \eqref{eq:vsensor}, the terms for the cantilever out of contact \eqref{eq:Qcl0}, and the heat-flows in contact \eqref{eq:Qcl}, \eqref{eq:Qts} into the energy equation in contact \eqref{eq:ICenergy}, then separating the equation for the constant and first and second harmonic contributions $n=1,2$, yields a system of three equations:
\begin{align}
    &\dVdc \cdot I = 
      \frac{\dVdc}{I\cdot \alpha \cdot \Rcl}
    + \frac{\dTsensorZero \cdot I \cdot \alpha + \dVdc}{I \cdot \alpha \cdot \Rts}
    - \frac{\dTdc}{\Rts},
    \label{eq_time_independet}\\
    &\dVacnw \cdot I = 
      \frac{\dVacnw }{I\cdot \alpha \cdot \Rcl} 
    + \frac{\dVacnw}{I\cdot \alpha \cdot \Rts} 
    - \frac{\betaN \cdot \dTdc}{\Rts}.
    \label{eq_time_dependet_nw}\\
    \intertext{The remaining unknown variables are the sample's temperature rise $\dTdc$ and the tip-sample thermal resistance $\Rts$.
    Equations \eqref{eq_time_independet} and \eqref{eq_time_dependet_nw}, together with \eqref{eq:const_current}, are used to solve for the unknown local sample temperature $\dTdc$ and the tip-sample thermal resistance $\Rts$, to finally infer the device's constant temperature rise to}
    &\boxed{\dTdc = \dTsensorZero \cdot \frac{\dVacnw }{\dVacnw - \betaN \cdot \dVdc }}. \label{T_DC_with_nw}\\
    \intertext{The harmonic $n=1,2$ can be chosen according to which signal has the larger signal-to-noise ratio, and the periodic temperature rise in the device $\dTacnw$ can then be calculated using \eqref{eq:beta_1} or \eqref{eq:beta_2} respectively. The tip-sample thermal resistance is calculated as}
    &\boxed{\Rts = \frac{[\dVdc +  ( \dTsensorZero - \dTdc ) \cdot I \cdot \alpha  ]\cdot \Rcl}{\dVdc \cdot [\RZero \cdot I^2 \cdot \alpha \cdot \Rcl - 1]}}\label{R_ts}
\end{align}
with $\Rcl = \dTsensorZero /( \VZero \cdot I)$.
    
In the case of no offset voltage applied to the device, $\Vdevdc =0$, we obtain $\beta_1=0$ and $\beta_2=1$, such that equation \eqref{T_DC_with_nw} simplifies to $\dTdc  =  \TsensorZero \cdot \dVacTwo /(\dVacTwo - \dVdc)$, as was found by Menges et al. \cite{Menges2016}. 

\begin{figure*}[t!]
\centering
\includegraphics[width=\textwidth]{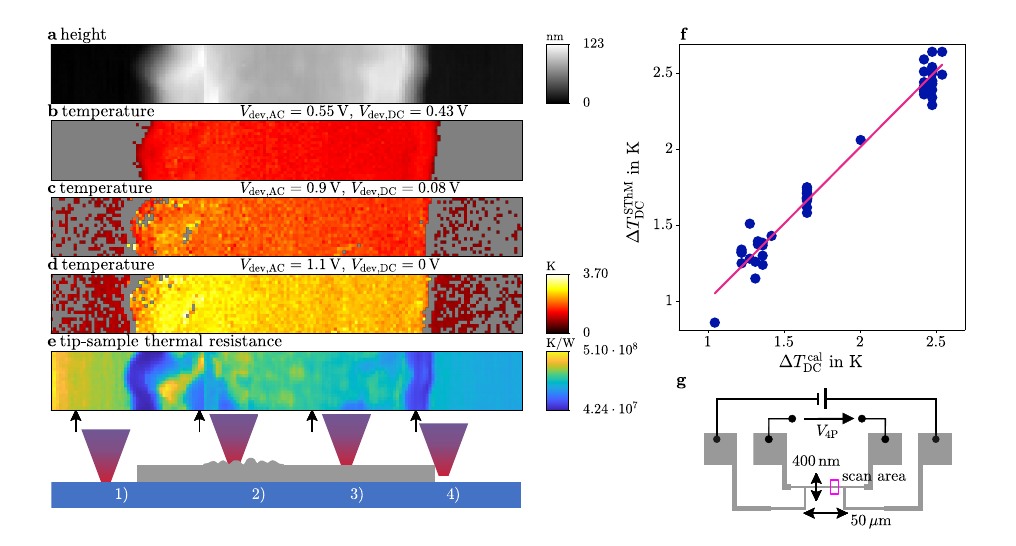}
\caption{Verification of the temperature measured in SThM. (a) Height measurement of line heater device (b)-(d) SThM temperature rise distribution at different DC and AC voltages applied. The temperature rise in each pixel (100x800 per image, with 5 nm resolution) is calculated from the equations for the non-constant current case, and the average temperature rise is calculated by taking the average over the device area in each image. (e) The tip-sample thermal resistance as measured in SThM, in different tip-sample contact cases: 1) sample region of low conductivity (substrate) 2) defective region of Pt film with large surface roughness 3) sample region Pt film of high conductivity 4) edge of Pt film with large contact area. (f) The averaged DC temperature rise measured in SThM, averaged per scan over the device area, plotted against the calibrated device temperature in blue and a linear least-squares fit $y = 1.0061\cdot x$ in pink. (g) Device schematic with the scan area indicated by the pink box.}
\label{fig:line_heater_SThM_quantitative_verification}
\end{figure*}

\subsection{Verification using a metal thin-film heater}
To verify that the proposed method is capable of measuring quantitative temperature rises, a metal thin-film heater device is imaged using this dual-scan SThM method. The temperature of the device is determined independently (see appendix). The device consists of a rectangular shaped Pt thin film of $50\:\micrometer$ length, $400$ nm width and $120\:\nanometer$ thickness with four contact pads deposited on a film of 80 nm $\SiO$ on a Si substrate. 
When operated as a self-heated device, its temperature rise in dependence of the applied current is calculated from the resistance change $\dTdccal = \Delta R/(\RdevZero \cdot TCR)$. Here the calibrated temperature coefficient of resistance of the Pt film is $TCR = 1.9\cdot 10^{-3} \,\mathrm{K^{-1}}$.

In a first step, the SThM sensor needs to be calibrated. As was identified by Spieser et al.\cite{Spieser2017}, the available fixed-point calibration can yield an error of up to $\sim 20\,\%$ for the empirical law describing the temperature of the sensor in relation to the electrical power of the cantilever, with an estimated maximum of $+10 \,\%$ error due to deviation of the specific cantilever used in this measurement. To improve the calibration, we use the already-verified dual-scan method without offset \cite{Menges2016}, i.e. only applying an AC voltage to the sample. A set of scans with different amplitudes $\Vdevac$ and no offset $\Vdevdc =0$ applied to the device were taken, and the temperature was found as described above. By comparing the expected with the measured temperature, the cantilever was calibrated, obtaining a working point of $\dTsensorZero=183\,\mathrm{K}$ and $\alpha=1.25\,\Omega/\mathrm{K}$. 

In a second step, we turn to the case of adding a large DC bias to the sample. The corresponding $\beta$-values are easily determined. For small $\omega$, we calculate:
\begin{align}
    &\dTdev \propto \Pdev = \frac{(\Vdevtot) ^2}{\Rdev}  \nonumber\\
    &\propto  \underbrace{(\Vdevdc)^2 + \frac{1}{2} (\Vdevac)^2}_{\propto \dTdc}
    \nonumber \\
    &\quad+ \underbrace{2 \, \Vdevdc  \cdot \Vdevac}_{\propto \dTacOne} \cdot \cos(\omega t)\nonumber \\
    &\quad + \underbrace{\frac{1}{2} (\Vdevac)^2 }_{\propto \dTacTwo} \cdot \cos(2 \omega  t). \label{eq:device_temp_oscillation}
    \intertext{which shows that the total temperature rise consists of a constant temperature rise and a periodic temperature rise in the first and second harmonic.}
    \intertext{Next, we calculate the ratios between the temperature rises
    }
    \beta_1 &= \frac{ \dTacOne}{\dTdc} = \frac{2 \, \Vdevdc \cdot \Vdevac  }{(\Vdevdc)^2 + \frac{1}{2} (\Vdevac)^2}  \label{eq:beta_1}
    \intertext{and}
    \beta_2 &= \frac{\dTacTwo}{\dTdc } = \frac{\frac{1}{2} (\Vdevac)^2 }{ (\Vdevdc)^2 + \frac{1}{2} (\Vdevdc)^2 }. \label{eq:beta_2}
\end{align}
Scanning Thermal Microscopy measurements were conducted for a range of device operating conditions that would result in a range of expected DC temperature rises $\dTdccal$ from $1\,\mathrm{K}$ to $2.5\,\mathrm{K}$ which were then compared to the SThM-measured temperature rises $\dTdcsthm$. Three examples of this set of measurements are shown in figure \ref{fig:line_heater_SThM_quantitative_verification}, where the temperature images in b-d illustrate the rise in temperature in the phase-locked pixels when raising the applied voltage. A detailed description of this plotting procedure is given in the appendix. We calculated the average temperature rise over the device area, as indicated in figure \ref{fig:line_heater_SThM_quantitative_verification} g by the pink box. The result of the measurement series is plotted in figure \ref{fig:line_heater_SThM_quantitative_verification} f, where each scan is represented by a single data point of its average DC-temperature rise. 
A line $y =m\cdot x$ was fit to all data points using a least-squares error function, yielding a slope $m=1.0061$ with goodness of fit $R^2=0.97$. This shows excellent agreement between the measured and calibrated device temperature using our new method. This result is a verification of the equations derived above, the experimental implementation, and, importantly, the underlying assumptions.

\section{Self heating of RRAM devices }

\begin{figure*}[t!]
    \centering
    \includegraphics[width =\textwidth]{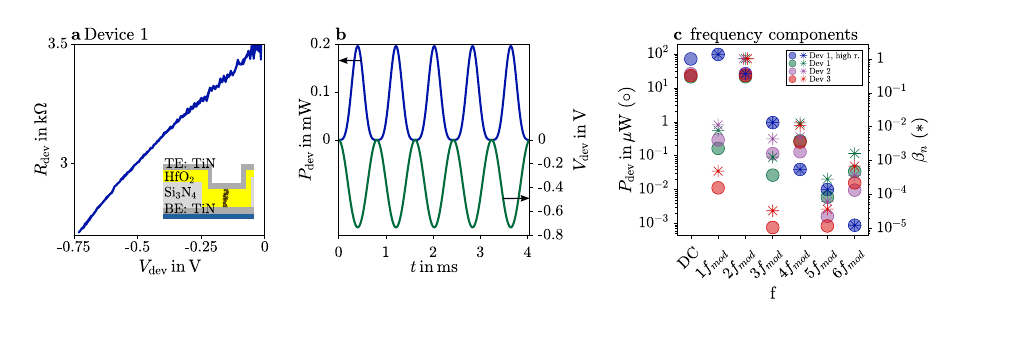}
    \caption{Resistance-Voltage characteristic of device 1 (a) and device voltage and power over time of device 1 in high resolution scan (b). Power frequency components, marked with filled circles, and resulting $\beta_n$, marked as asterisks, of device 1 for high (blue) and low (green) resolution images and device 2 (lilac) and 3 (red) (c). }
    \label{fig:Nonlineardevice_RV_betas}
\end{figure*}

\begin{figure*}[t!]
\centering
\includegraphics[width=\textwidth]{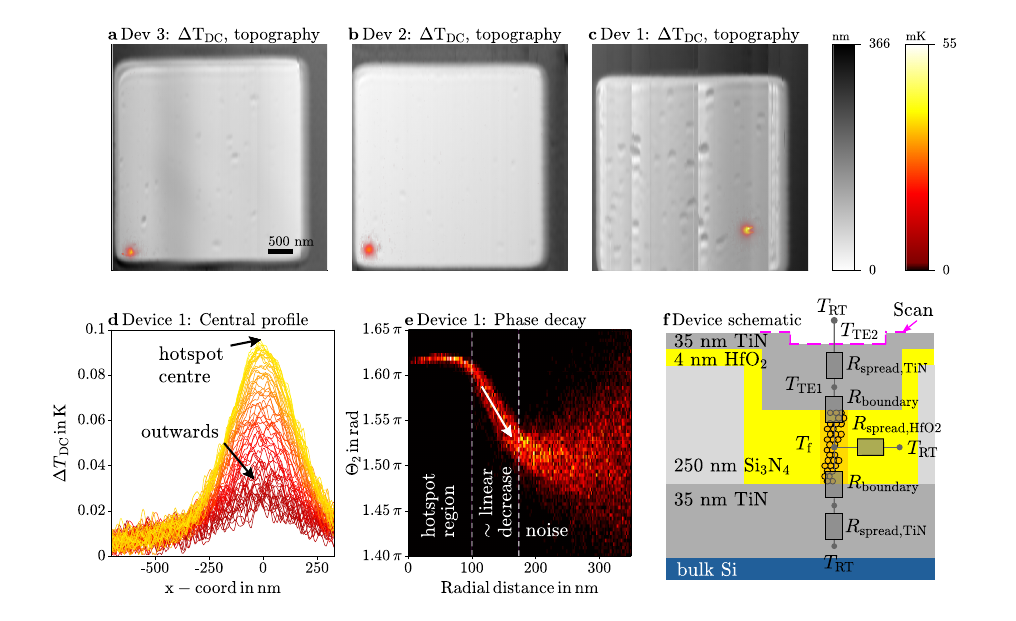}
\caption{SThM temperature images of three different RRAM devices without offset applied (a-c). DC temperature profiles through center of hotspot (d) and decrease of phase with radial distance to hotspot center (e) from high resolution scan of device 1 with offset applied. The full data and detailed analysis including the detection of an artifacts in the hotspot region of this scan without offset applied is presented in the appendix. Device schematic with thermal resistances (f).}
\label{fig:Nonlinear_device_SThM_images}
\end{figure*}

\subsection{$\mathrm{HfO_2}$-based RRAM devices: open questions}

Hafnium Oxide ($\mathrm{HfO_2}$)-based memristors have been studied as candidates for neuromorphic computing devices \cite{Dittmann2021}. To operate devices in a bipolar way, applying a voltage $V_{\mathrm{SET}}$ to the device is used to switch the device into the low resistive state (LRS), and a voltage $V_{\mathrm{RESET}}$ of opposite polarity can be applied to reset it to the high resistive state (HRS). The fundamental mechanisms, however, are still being explored. Nanoionic processes, i.e. the migration of ions/oxygen vacancies, are used to explain the switching and failure mechanisms in $\mathrm{HfO_2}$-based and other RRAM devices \cite{Dittmann2021}. For example during SET switching, the migration of ions is dependent on temperature. This creates a positive feedback loop via the increased Joule heating with decreasing resistance, and accelerates the closing of the ruptured filament's gap to retain the LRS. The dependence on temperature is often described and parametrised as a kinetic process, which explains the dependence of retention time, switching time and filament size on compliance current to a satisfactory level.\cite{Raffone2018, Wei2011, Celano2015} 

However, the actual magnitude of the temperature inside devices is often unknown. Moreover, the heat dissipation in highly-integrated crossbar arrays needs to be carefully engineered to avoid thermal crosstalk between neighbouring devices, which lower retention time and influence the robustness of operation parameters. However, the actual size of the heat source, which is not necessarily equal to the filament, its temperature, which is not necessarily uniform, and the decay of temperature in the different layers depend on many factors. They are determined by the geometry, mode of transport (diffusive/ballistic/quantum), thermal conductances, and thermal boundary resistances of the surrounding materials and interfaces.
Therefore, a deeper understanding of the thermal processes inside RRAM, lead by highly resolved thermometry, is essential for reliability and ultimate scaling goals. \\

To address these open questions using the method introduced here, we studied $\mathrm{HfO_2}$-based RRAM fabricated and electrically characterised by Stellari et al. \cite{Stellari2021}. The device design is schematically depicted in figures \ref{fig:first_illustration_thermal_resistances} c and \ref{fig:Nonlineardevice_RV_betas} a. It consists of an active layer of 4\,nm $\mathrm{HfO_2}$ with a 35\,nm TiN bottom- and top electrode on a bulk Si substrate. The separation between the electrodes is achieved by a $\mathrm{Si_3N_4}$ interlayer, in which an opening of $4\,\mu \mathrm{m}$ x $4\,\mu \mathrm{m}$ defines the device area.

\subsection{Dual scan SThM method applied to non-linear devices}

In the case of a non-linear device $R_{\mathrm{dev}} \neq $const with Joule heating, a numerical solution for $\beta_1$ and $\beta_2$ can be calculated using the resistance-voltage characteristics. Subsequently, equations \eqref{T_DC_with_nw}, \eqref{R_ts} can be used to infer the temperature and the tip-sample thermal resistance as before.

To find the temperature ratios $\beta_1$ and $\beta_2$, we need the temperature rise over time, which is again assumed to be proportional to the dissipated heat over time, $\Delta T(t) \propto \Pdev(t)$. We start with the device resistance-voltage relationship as shown in figure \ref{fig:Nonlineardevice_RV_betas} a, and calculate the dissipated device power as a function of voltage $\Pdev(\Vdevtot)$. Then we insert the time dependence, equation (\ref{eq_time_dependet_nw}), to calculate $\Pdev(t)$, as shown in figure \ref{fig:Nonlineardevice_RV_betas} b. Thirdly, a Discrete-Fourier-Transform is applied to $\Pdev(t)$ to determine the contributions of the harmonics of the modulation frequency, the result of which is shown in figure \ref{fig:Nonlineardevice_RV_betas} c with filled circles. Next, the amplitudes of the constant part $\Pdevdc= 71\,\mathrm{\mu W}$ and the oscillations at the modulation frequency $\PdevacOne = 97\,\mathrm{\mu W}$ and double the modulation frequency $\PdevacTwo = 27\, \mathrm{\mu W}$ are extracted. The $\beta$ are calculated to $\beta_1 = \PdevacOne / \Pdevdc = 1.35$ and $\beta_2 =  \PdevacTwo / \Pdevdc  = 0.37$. The calculated $\betaN$ are shown in \ref{fig:Nonlineardevice_RV_betas} c as asterisks for all higher harmonics. The values rapidly decrease with increasing $n$ and have high values for $n=2$, justifying basing the analysis on this harmonic. 
Finally, equation \eqref{T_DC_with_nw} is used to infer the local constant device temperature rise $\dTdc$ from the measured $\dVacTwo$ and $\dVdc$. 

\subsection{Results and discussion}
The resulting DC temperature rises are shown in figure 
\ref{fig:Nonlinear_device_SThM_images} for three different devices. In each of the devices, only a single hotspot is observed, which we assign to single current filaments from which heat spreads into the top electrode. In addition, the position of the hotspot does not appear to vary during the course of our measurement series, i.e. operation over several hours. An important feedback possible due to the high resolution is that the position of the filament is not located at the device edges, where it might form due to compositional fluctuations and defects. 

{\bf Hot-spot diameter:} To discuss the diameter of the observed hotspot, an additional scan of only the hotspot was recorded for device 1 with a higher lateral resolution of $5\,\mathrm{nm}$, of which the temperature profiles through the center and at different offsets are shown in \ref{fig:Nonlinear_device_SThM_images} d. The temperature reaches about $0.1\,\mathrm{K}$ in the center of the hotspot with a narrow concave region around the center followed by a broad decay up to a radius of at least $200\,\mathrm{nm}
$. 

To find the radius of the underlying heat-source, the heat spreading in the top electrode can be described by spreading from an isothermal circular heat-source \cite{Yovanovich2006} (for large top electrode thickness $\tTiN$, see supplementary). Alternatively, the radius can be found by calculating the inflection point of the temperature profile, which was confirmed by finite element simulation to overestimate the radius by only $\sim$10 nm in the system at hand. The extracted heat-source radius using the shape of the temperature decay is $\rf \approx 100-133\,\mathrm{nm}$. We note here that the extracted quantity is the heat-source radius, and not the filament radius. Still, this appears large compared to the expected diameter of the conductive filament of nanoscale down to molecular dimension. However, the exact biasing procedure and material stack can influence the size \cite{Dittmann2021}. 
In addition, heat spreading takes place inside the $\mathrm{HfO_2}$ layer, which could lead to further broadening of the effectively seen heat source. The heat-spreading could even lower the electrical conductivity in its surrounding, also leading to a larger heat-source diameter \cite{Kwon2016}.

{\bf Hot-spot temperature:} Turning to the magnitude of the temperature rise, we note that the temperature rise is much smaller than the expected temperature of the current filament during switching on the order of hundreds of Kelvin. As already discussed in previous work \cite{Deshmukh2022,Swoboda2023,West2023}, one explanation is the large thermal conductivity and heat-spreading in the metal top electrode. However, we find from finite-element simulations of the top electrode only, that the maximum temperature reached at the surface is very similar to the temperature at the electrode side of the material interface. Moreover, as pointed out by Deshmukh et al. \cite{Deshmukh2022}, even more important is the thermal boundary resistance between the oxide and the metal electrode. At this point it may be most instructive to discuss their influence using approximate analytical equations that show the expected scaling with dimension. A simplified thermal circuit is depicted in figure \ref{fig:Nonlinear_device_SThM_images} f. Starting with the thermal boundary resistance \cite{Chen2005} between TiN and $\mathrm{HfO_2}$
\begin{align}\label{EqSpread}
    \Rboundary = \left( h_{\mathrm{k}} \cdot \pi \cdot r_{\mathrm{f}} ^2 \right) ^{-1}
\end{align}
we use $\hk = 1\,\mathrm{MW\cdot m^{-2}K^{-1}}$, and obtain about $\Rboundary \approx 3.2 \cdot 10^7\,\mathrm{K/W}$. Next, the heat spreads into the top electrode, which for $\tTiN \gg \rf$ one may approximate by 
\begin{align}
    \RspreadTiN =\frac{1}{4 \cdot \rf \cdot \kappaTiN}
\end{align}
Using $\kappaTiN \ \approx 11$\,W/(mK) \cite{Samani2015} we obtain $\RspreadTiN \approx 2.2\cdot10^5$\,K/W. However, in the case at hand the filament diameter is larger than the thickness of the film. Therefore we use a finite element simulation to calculate this resistance, resulting in $\RspreadTiN =1.7\cdot10^6$\,K/W, which is larger by one order of magnitude. Nevertheless, the boundary resistance is larger than the spreading resistance by one order of magnitude, with this ratio becoming even more pronounced if smaller filaments are expected.
To understand the small magnitude of temperature drop at the electrode surface, we have to compare these two resistances to the further heat spreading inside the $\mathrm{HfO_2}$.
The cylindrical heat spreading equation predicts a logarithmic scaling of thermal resistance with the size of the filament :
\begin{align}
\RspreadHfOTwo= \frac{\ln \left( \frac{\rspreadHfOTwo}{\rf} \right)}{2 \cdot \pi \cdot \tHfOTwo \cdot \kappaHfOTwo}
\end{align}
with a thermal conductivity of $\mathrm{HfO_2}$ on the order of $\kappaHfOTwo \approx 1$\,W/(mK), a thickness of $\tHfOTwo = 4$\,nm and a decay to room temperature at radius $\rspreadHfOTwo\approx \,100\,\mathrm{nm}$, we obtain about $\RspreadHfOTwo \approx 2.6\cdot10^5$\,K/W. This is two orders of magnitude smaller than the thermal boundary resistance, and one order of magnitude smaller than the spreading resistance in the TiN layer.
The total thermal resistance and temperature are calculated to $\Rtot =2.56\cdot10^5\,\mathrm{K/W}$ and, using the DC electrical power calculated above, $\Delta \Tf=18.2\,\mathrm{K}$ above room temperature. This is a plausible temperature since during our measurement, which is equivalent to a read operation, the filament is stable and does not switch. These conclusions are similar in terms of hotspot size and in support of previous findings \cite{Deshmukh2022,Swoboda2023,West2023},

{\bf Noise discrimination:} The method presented in this work provides further insight: there is additional useful information in the phase signal obtained in the phase-sensitive detection of the thermal signal using a lock-in amplifier. If there is a clear phase signal with a value indicative of phase-locking, then we can confidently analyse the results. In return, a noise-phase signal is indicative of a lack of phase-locking and we can conclude that the thermal amplitude signal is not related to the device heating by the applied AC current. This is an important and useful way to discriminate errors and confidently tell apart small signals from noise, which appears more difficult to do using conventional SThM techniques. The temperature signal in figure \ref{fig:Nonlinear_device_SThM_images} is plotted in such a way that only in the significant (phase locked) regions a temperature is indicated by the colour bar, while other regions remain in the greyscale of the topography image. We found this a useful way to display data for various samples. More details are found in the Supplementary information. 

{\bf Observation of diffusive wave:} The phase information has even more merits. We can identify regions in which the phase is locked to the electrical excitation at a different value from the one in the center of the hotspot. In such regions, a phase lag can be attributed to a time delay in heat spreading. In figure \ref{fig:Nonlinear_device_SThM_images} e we report the radial evolution of the phase signal away from the center. Three different regions can be identified. First, in a radius of up to $100\,\mathrm{nm}$ around the hotspot center, the phase is almost constant. In this region, the ratio of signal amplitudes is also constant and close to the expected ratio $\beta_1/\beta_2$ (see supplementary). This confirms the assumption of the temperature following the power without time delay, i.e. steady state operation. The region without phase locking is seen further out for radial distances larger than about 180 nm. As stated above, the associated small but still visible thermal amplitude signal needs to be handled with care.

In between these trivial regions, we observe an almost linear decrease of the signal with increasing radius. In this region, we attribute a phase lag to the finite thermal diffusivity. Importantly and excitingly we can access this observable not easily obtained for nanoscale materials. We intend to extract quantitative values of diffusivity from such data in a dedicated study. A phase deviation can generally be used to avoid errors in making a steady state assumption. We give an example in the Supporting information.

{\bf Device design considerations:} 
To motivate further study, we draw several preliminary conclusions from these observations. 

According to our calculations, the total thermal resistance and temperature of the filament are dominated by the spreading inside the $\mathrm{HfO_2}$, which is made possible by the large thermal boundary resistance. Moreover, the small observable temperature at the top-electrode is dominated by the large thermal boundary resistance, which confines the heat to dissipate through the $\mathrm{HfO_2}$ layer.
In combination with modelling, which takes the geometries into account, the SThM data feeds into thermal conductivity, thermal diffusivity and thermal boundary conductance values. This is important because these quantities are dependent on the deposition process and film thickness. Therefore, relying on literature values may lead to a significant systematic error in modelling.

\section{Conclusion}
In this paper we have derived a method to obtain in-operando quantitative temperature information about nanoscale non-linear devices. We verified the proposed approach by imaging a calibrated device of homogeneous temperature and found good agreement between the measured and expected temperatures, if the sensor's temperature is well known and calibrated. The used approach can be applied easily to a network of non-linear devices if the $\beta_1$ and $\beta_2$ of the network memristive elements are equal; else it has to be extended further. We also show how the temperature rise due to Joule and Peltier heating can be distinguished using this method. The proposed method mitigates artifacts from tip-sample thermal resistance variations. Additionally, phase information can be used to identify regions where the calculated temperature and tip-sample thermal resistance are still erroneous, as is shown in images of RRAM devices.

The method was applied to $\mathrm{HfO_2}$-RRAM devices. The low temperature values suggest that the thermal boundary resistance cannot be ignored in such devices, especially with smaller radii of heat-sources than found in our devices. Important information (effective heat-source radius) can be extracted from high-resolution temperature images, especially when other information is missing.

\subsection{Acknowledgements}
We thank Beatriz Noheda and Achim Kittel for their generous support through academic supervision of N.H and S.R., respectively. We thank Fabian Könemann for his work on the SThM method and setup on which this project is based.\\
We also thank Franco Stellari and Takashi Ando for providing the samples and guidance.
Continuous support from and discussions with Siegfried Karg, Olivier Maher, and Fabian Menges are gratefully acknowledged.
We thank the former Materials Integration Nanoscale Devices (MIND) group, the Physics and Science of Information group and the Cleanroom Operations Team of the Binnig and Rohrer Nanotechnology Center (BRNC) at IBM Research Europe Zurich for their help and support. \\
This project has received funding from the EU’s Horizon 2020 programme under the Marie Skłodowska-Curie grant agreement No 861153, and from the SNSF project HYDRONICS under the Sinergia Grant 315 (No. 189924).

\section{Appendix}

\subsection{Derivation of equations for non-constant current} 
So far, we have derived equation \ref{T_DC_with_nw} to calculate the sample's temperature rise for the case of constant current in the sensor. This holds true as long as the resistance changes in the cantilever are negligibly small in comparison to the constant resistance in the Wheatstone-bridge.  This can be enhanced, at the cost of lower sensibility to sensor-resistance changes, by choosing a series resistor in the Wheatstone-bridge which is much larger than the sensor's resistance.\\ In the general case of $I \neq const.$, we need to rewrite equation \ref{eq:const_current} as 
\begin{align}
I(t) &= \frac{\Vtot}{\Rseries+\Rsensor(t)}\label{eq:non_const_current}
\intertext{which leads to the expression}
\Rsensor(t) &= \frac{\Vsensor(t) \cdot \Rseries}{\Vtot - \Vsensor(t)}
\end{align}
which is not trivial to split into the different constant and frequency components due to $\Vsensor(t)$ appearing in both nominator and denominator, but is necessary to obtain the sensor's resistances $\dRdc$, $\dRacOne$ and $\dRacTwo$ with their corresponding temperature changes and the electrical power components $\Peldc$, $\dPelacOne$ and $\dPelacTwo$.

We note that in the Wheatstone-bridge, when the cantilever's resistance changes periodically at a frequency $f_0$, the Wheatstone-bridge voltage will have a component at that frequency and, due to the non-constant current, also components at higher harmonics of the frequency with decreasing amplitude. To describe the signal created in higher harmonics, we can write the Wheatstone-bridge voltage dependent on the (unknown) cantilever resistance change $\dRacOne$ at angular frequency $\omega_0=2\pi f_0$ as

\begin{align}
\frac{\Vsensor(t)}{\Vtot} =& \frac{\VZero + \dV}{\Vtot} \nonumber \\
=& \frac{\RZero + \dRdc + \dRacOne \cdot \cos(\omega_0 t)}{\Rseries + \RZero + \dRdc + \dRacOne \cdot \cos(\omega_0 t)} \nonumber \\
=& \frac{\epsilon + \cos(\omega_0 t)}{\gamma + \cos(\omega_0 t)}
\intertext{where $\epsilon \coloneqq (\RZero + \dRdc)/\dRacOne$ and $\gamma \coloneqq (\Rseries + \RZero + \dRdc)/\dRacOne$. Unfortunately, there exists no analytical solution for the Fourier transform of a signal of this form}
\frac{\Vsensor(\omega)}{\Vtot} =&  \int_{-\infty}^{\infty} \frac{\epsilon + \cos(\omega_0  t)}{\gamma + \cos(\omega_0 t)} \cdot e^{-i \omega t} dt
\intertext{Instead, we obtain the coefficients of the Fourier series of $\Vsensor(t)$ by means of comparison of coefficients:}
\frac{\epsilon + \cos(\omega_0 t)}{\gamma + \cos(\omega_0 t)} =& A_0 + \sum_{n=1}^{\infty} \left( \An \cdot \cos(n \omega_0 t) + \Bn \cdot \sin(n \omega_0  t) \right) \nonumber
\end{align}

\begin{align}
&\epsilon + \cos(\omega_0  t) = A_0 \cdot [\cos(\omega_0 t) + \gamma] 
\nonumber \\ &+ \sum_{n=1}^{\infty}  \An \cdot \cos(n \omega_0 t)\cdot [\cos(\omega_0 t) + \gamma]\nonumber\\
&\, + \Bn \cdot \cos(n \omega_0 t)\cdot [\cos(\omega_0 t) + \gamma]\nonumber\\
=& A_0 \cdot [\cos(\omega_0 t) + \gamma] \nonumber\\
+& \sum_{n=1}^{\infty} \An \cdot [\onehalf \cos((n-1) \omega_0 t) + \gamma \cdot \cos(n \omega_0 t) + \onehalf \cos((n+1) \omega_0 t)]  \nonumber\\
+& \sum_{n=1}^{\infty} \Bn \cdot [\onehalf \sin((n-1) \omega_0 t) + \gamma \cdot \sin(n \omega_0 t) + \onehalf \sin((n+1) \omega_0 t)]  \nonumber
\end{align}

Comparing the coefficients, we see that all $\Bn = 0$ and are left with $n$ equations for $n+1$ coefficients $\An$. Since we know that the $\An$ decrease monotonically, we can reduce the number of variables to $n$ by setting $\Ainf = 0$ and can solve for an expression for the $\An$ that only depends on the $x_n$

\begin{align}
&\An = [(-1)^{n+1} \cdot 2\gamma + (-1)^{n} \cdot 2\epsilon] \cdot [\frac{1}{x_{n+1}} +x_n \sum_{i=n+1}^\infty \frac{1}{x_i \cdot} x_{i+1}] \nonumber
\intertext{with the $x_n$}
&x_0 \coloneqq \onehalf \nonumber \\
&x_1 \coloneqq \gamma \nonumber \\
&x_2 \coloneqq 2\gamma^2 -1 \nonumber \\
&x_n \coloneqq 2\gamma x_{n-1} - x_{n-2} \text{ for } n \geq 3 \nonumber
\intertext{and find the ratio between the signal at harmonic $n$ and the following harmonic $n+1$}
&\frac{A_{n+1}}{A_n} = (-1) \frac{[\frac{1}{x_{n+2}} +x_{n+1} \sum_{i=n+2}^\infty \frac{1}{x_i \cdot} x_{i+1}]}{[\frac{1}{x_{n+1}} +x_n \sum_{i=n+1}^\infty \frac{1}{x_i \cdot} x_{i+1}]}
\nonumber \\
\intertext{using $\gamma \gg 1$ for the typical values during the measurement, this finally yields the ratio}
&\frac{A_{n+1}}{A_n} = -\frac{1}{2\gamma}
\intertext{We can interpret the result as the amplitude of V becoming smaller with $\frac{1}{2\gamma}$ while the alternating sign of the amplitude corresponds to a phase shift of $\pi$.
In our case, $\gamma$ is large, and when detecting the resistance change in $1 \omega$ it is sufficient to use the voltage signal detected in $1 \omega$, neglecting the $2 \omega$ signal.
However, when sufficiently high $1 \omega$ resistance change is present, it creates a $2 \omega$ component in the voltage as well, so that the $2 \omega$ voltage signal can not readily be used to calculate the $2 \omega$ resistance change. 
In such cases, it might be beneficial to replace the voltage-source with a current source.}
\intertext{In the experiment, we can either use a Taylor-expansion, here to the second order, to calculate the sensor's resistance}
R &= \frac{\VZero}{\IZero} + \frac{\Vtot \cdot \Delta V}{\IZero \cdot (\Vtot - \VZero)} + \frac{\Vtot \cdot \Delta V^2}{\IZero \cdot (\Vtot - V_0)^2} \label{eq:R non-const I} 
\intertext{or a Discrete-Fourier-Transform on the time dependent signal $\Rsensor(t)$.
The electrical power in the heater can then be calculated to and separated into three different contributions at $1\omega$, $2\omega$ and DC:}
\Pel(t) &= \Vsensor(t) \cdot \frac{\Vtot - \Vsensor(t)}{\Rseries} \nonumber \\
&= \frac{(\VZero + \dVdc)(\Vtot - \VZero - \dVdc)}{\Rseries}\nonumber \\&+ \frac{\dVacOne(\Vtot -2\VZero - 2\dVdc)}{\Rseries} \cdot \cos(\omega t) \nonumber \\ &+  \frac{\dVacTwo(\Vtot -2\VZero - 2\dVdc)}{\Rseries}  \cdot \cos(2 \omega t) \label{eq:Pel non-const I}
\end{align}
using \eqref{eq:vsensor} and neglecting second order terms $\dVacOne \cdot \dVacTwo$, $\dVacOne^2$ and $\dVacTwo^2$. Now we can use the obtained $\dRdc$, $\dRac$, $\Peldc$ and $\dPelac$ together with the initial equations \eqref{eq:Qcl0} to \eqref{eq:Qts} to solve for the desired sample temperature and tip-sample thermal resistance:
\[\boxed{
\!\begin{aligned}
&\dTdc \nonumber \\
&= \frac{\Rcl \cdot [ \dRac  \Peldc - ( \dTsensorZero \alpha + \dRdc ) \dPelac ] }{\Rcl  \alpha (\beta \Peldc -\dPelac ) - (\dTsensorZero \alpha + \dRdc) + \dRac}
\end{aligned}
}\]
\begin{align} \label{eq:dTdc non-const I} 
\end{align}
and

\begin{align}
&\boxed{\Rts = \frac{\Rcl \cdot ( \dTsensorZero + \dTsensordc - \dTdc )}{ \Rcl \cdot \Peldc - (\dTsensorZero + \dTsensordc ) }}. \label{eq:Rts non-const I}
\end{align}
To ease reading, the results of \eqref{eq:dTdc non-const I} were not plugged into \eqref{eq:Rts non-const I}.

\subsection{Combining Joule heating with Peltier effects}

To take into account and quantify the Peltier effect relevant for some device types, the simultaneous measurement at different harmonics can be used. Because the Peltier cooling/heating scales with the current $I$ while the Joule effect scales with power (or $I^2$ for linear devices) the two effects appear differently at the different harmonics. In the case shown above, comparing the different harmonics confirms that Peltier effects can be neglected. However, to indicate how to proceed in other cases, we show the equations for a linear device below, which also shows how to extend the calculations to non-linear cases.\newline
In the case of a linear device that shows Joule heating and Peltier heating/cooling, the temperature rise in the device can be inferred as
\begin{align}
    &\dTdev \propto \Pdev = \frac{\Vdevtot^2}{\Rdev} + \Pi \cdot I = \frac{\Vdevtot^2}{\Rdev} + \Pi \cdot \frac{\Vdevtot}{\Rdev} \nonumber\\
    &\propto  \underbrace{\Vdevdc^2 + \frac{1}{2} \Vdevac^2 + \Pi \cdot \Vdevdc }_{\propto \dTdc}\nonumber\\
    &+ \underbrace{(2 \, \Vdevdc \cdot \Vdevac  + \Pi \cdot \Vdevac )}_{\propto \dTacOne} \cdot \cos(\omega t)\nonumber \\
    &+ \underbrace{\frac{1}{2} \Vdevac^2}_{\propto \dTacTwo} \cdot \cos(2 \omega  t) \label{eq:device_temp_oscillation_w_Peltier}
    \intertext{where $\Pi$ is the Peltier-coefficient. This takes into account the temperature change due to Peltier heating/cooling, but does not correspond to the local thermovoltage directly due to heat propagation in the device. The definitions of $\beta_1$ and $\beta_2$ area used according to \eqref{eq:beta_1} and \eqref{eq:beta_2}, as before only for the Joule-heating temperature rises. Then the initial equations \eqref{eq:Qcl0} to \eqref{eq:Qts} are used again and separated into their constant, $1\omega$- and $2\omega$-components. We obtain three equations which implicitely contain the variables $\dTdc$, $\Rts$ and $\Pi$ such that all equations have to be used in the case at hand. Using the frequency components of the sensor’s resistance $\dRdc$, $\dRacOne$, $\dRacTwo$ and electrical power $\Peldc$, $\dPelacOne$, $\dPelacTwo$ obtained with the constant current or alternative method as described above, the sample temperature can be calculated:}
    &\dTdc =  \nonumber \\
    &\frac{\Rcl \cdot [ \dRacTwo  \Peldc - ( \dTsensorZero \alpha + \dRdc ) \dPelacTwo ] }{\Rcl  \alpha (\beta_2 \Peldc -\dPelacTwo ) - \beta_2 (\dTsensorZero \alpha + \dRdc) + \dRacTwo}. \label{eq:dTdc non-const I w Peltier}
    \intertext{We note that the $2\omega$-component must be used in his case and is not interchangeable with the $1\omega$-component. The tip-sample resistance can be calculated using the previously derived equation \eqref{eq:Rts non-const I}, again not plugging in \eqref{eq:dTdc non-const I w Peltier} for readability. The Peltier-coefficient is calculated to}
    &\Pi = \nonumber \\
    &\frac{g_2 \cdot (\dRacOne -\beta_1 \alpha \dTdc) - g_1 \cdot (\dRacOne - \beta_2 \alpha \dTdc)}{\Vdevac \alpha g_2}
    \intertext{with $g_1=\dPelacOne \alpha \Rcl - \dRacOne$ and $g_2=\dPelacTwo \alpha \Rcl - \dRacTwo$. The constant and first harmonic temperature rises in the device due to the Peltier effect can then be calculated from}
    &\dTdcPeltier = \dTdc \cdot \frac{\Pi \cdot \Vdevdc }{ \Vdevdc^2 + \frac{1}{2} \Vdevac^2+ \Pi \cdot \Vdevdc},\\
    &\dTacPeltier = \frac{\dTdc \beta_1 \Pi}{2 \Vdevdc}
\end{align}
and the temperature rise due to Joule heating can be found using \eqref{eq:beta_1} and \eqref{eq:beta_2}.

\newpage
\section{Supplementary}

\subsection{Experimental Setup}

The Scanning Thermal Microscopy was conducted in the setup at the BRNC noise free labs in Zurich, Switzerland. The labs provide shielding against electromagnetic, acoustic and vibrational disturbances and are temperature and humidity stabilised, which is described further in Lörtscher et al. \cite{Lortscher2013}. The home built microscope described in more detail in Menges et al. \cite{Menges2016} is set up in a high vacuum chamber at a pressure of $10^{-7}\,\rm{mbar}$ and at a temperature of $21\:^{\circ}\rm{C}$. The thermoresistive probe utilising a doped Si-heater used here is biased in a Wheatstone-bridge with 2.5 V, which conforms to $P=1.3\,\rm{mW}$. It is calibrated prior to the measurement and consists of the sensor temperature at the working point $\TsensorZero$ and its change of resistance with temperature change $\alpha$, where $\dTsensorZero$ contributes linearly to the measured temperature as can be seen from \eqref{T_DC_with_nw}. The full procedure is described in Spieser et al. \cite{Spieser2017}. We note here that the calibration in this approach does not include the tip-sample thermal resistance, but solely describes the probe: the thermal resistance of the cantilever legs, $\Rcl$, and the temperature of the sensor out of contact with the sample, $\TsensorZero$.

To conduct the SThM measurements of the $\mathrm{HfO_2}$ devices, the chip with the devices was glued to a printed circuit board using silver paint, which at the same time provides an electrical contact for the bottom electrode. The top electrode contact was wire bonded with $25\,\mathrm{\mu m}$ Al wire in a Devoltek wire bonder. The devices show binary switching between a high and a low resistive state HRS and LRS.

\subsection{Details of the metal thin film heater (calibration sample)}

Due to the very high aspect ratio and high thermal conductivity within the metal heater device, the resulting temperature distribution is expected to be homogeneous along its length. Therefore a small part of the total surface area, such as in a scan, is representative of the overall temperature. Moreover, the average device temperature can be determined through measuring its resistance change in $4$-probe configuration as a function of the applied current.

To calibrate the sample, it was placed in the temperature controlled environment of a DynaCool PPMS. By applying a temperature ramp and reading out the 4P-resistance with a voltage small enough not to cause self-heating, the resistance-temperature relation of the device was measured.

\subsection{Laser heating in cantilever}

The cantilever deflection which is used for the AFM-capability of SThM is measured via the displacement of the laser dot reflected from the cantilever's top, as is typically done in AFM systems. In this case we also need to take into account that the light is not fully reflected from the cantilever. We need to consider that $\Pel$ is not the only 'input' heat flow during the calibration of the sensor both during the calibration and the measurement by replacing $\Pel$ with $\Pel + \Plaser$ in the energy equation.
For calibration of the cantilever, we can see a shift in power between the R-P curve taken with or without the laser focused on the cantilever. By calculating this shift, we can determine the laser heating. During the measurement, we can replace the DC-energy equation by $\Pel + \Plaser = \Qcldc + \Qtsdc$ while the AC-energy equation stays unaffected. 
In our setup, the contribution from the laser typically is a very small correction.

\subsection{Phase for artifact detection and full data set: resistive RAM}

In the main manuscript, we explain the usefulness of detecting the phase relationship of the thermal signal with respect to the applied device voltage. Here we give two examples, showing how the phase signal is used to verify that the assumptions of local equilibrium and no interference from the measurement process are valid. In addition, we give more consideration to the thermal wave interpretation.
 
\begin{figure*}
\centering
\includegraphics[width=\textwidth]{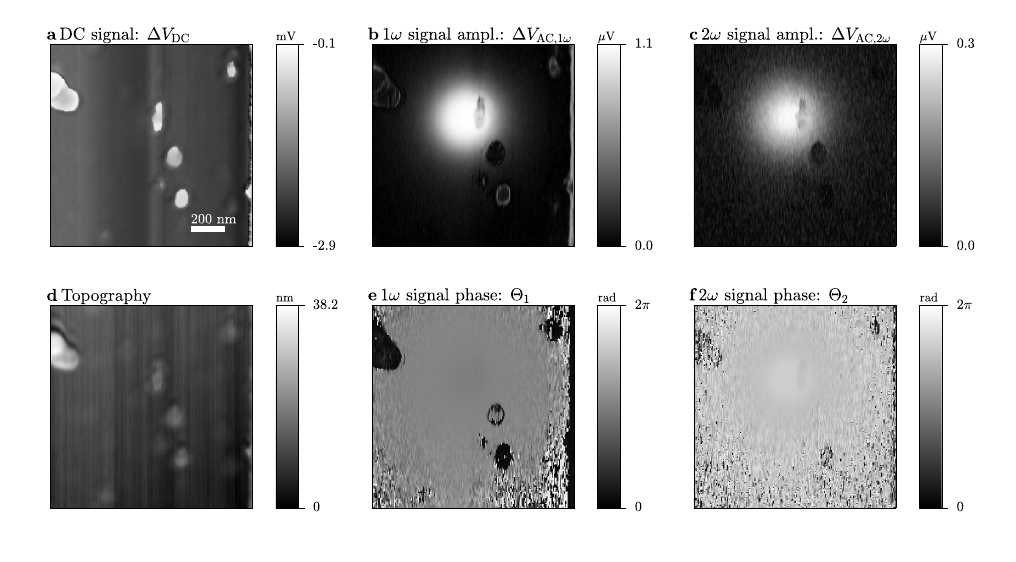}
\caption{High resolution (5 nm) image of hotspot in device 1. DC voltage change relative to out-of-contact value (a) and AC voltage amplitude in $1\omega$ (b) and $2\omega$ (c). Topography of device (d) and AC voltage phases in $1\omega$ (e) and $2\omega$ (f). The $1\omega$ signal is corrected for crosstalk: the amplitude $\Delta V_{\mathrm{cross}}$ and phase $\Theta_{\mathrm{cross}}$ of electrical crosstalk are extracted from 30x30 pixels in the bottom left corner of the raw signal. The $1\omega$ signal is then corrected to the complex $\dVacOneC \cdot \exp^{i\cdot\Theta_{1,\mathrm{C}}}- \Delta V_{\mathrm{cross}}\cdot \exp^{i\cdot\Theta_{\mathrm{cross}}}$, of which the amplitude and phase are plotted in (b) and (e).}
\label{fig:Nonlinear_device_SThM_raw_data}
\end{figure*}

\begin{figure}
\centering
\includegraphics[width=0.5\textwidth]{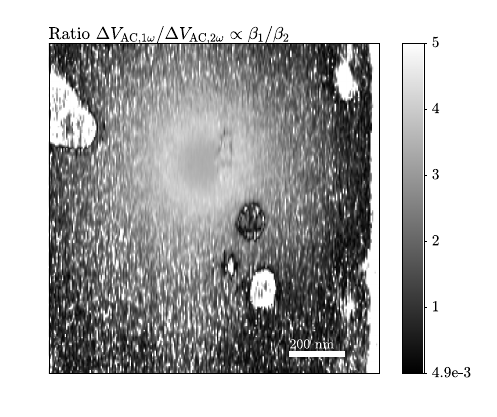}
\caption{Ratio of the amplitude signals $\dVacOne/\dVacTwo$ with measured central value of $~3.4$, which is slightly smaller but consistent with the expected value $\beta_1/\beta_2$=3.7. It increases with the radial distance, indicative of amplitude decay in the thermal wave.}
\label{fig:Nonlinear_device_SThM_beta_ratio}
\end{figure}

In figure \ref{fig:Nonlinear_device_SThM_raw_data} the full dataset of the scan of device 1 is shown, from which the line scans in figure \ref{fig:Nonlinear_device_SThM_images} d were extracted. The hotspot region is visible in both amplitude (region of largest amplitude) and phase signals (region with constant phase), where each signal decreases monotonically with increasing radial distance from the center of the hotspot. The amplitude of the temperature decreases due to heat spreading, while the decrease in phase can be explained by the presence of thermal waves. The presence of thermal waves can be confirmed by taking the ratio of the values of $\dVacOne$ and $\dVacTwo$ as depicted in figure \ref{fig:Nonlinear_device_SThM_beta_ratio}. They are consistent with the ratio of the predicted $\beta_1$ and $\beta_2$: $\dVacOne/\dVacTwo\approx3.4$ and $\beta_1/\beta_2\approx3.7$ in the center of the scan. They increase with increasing radial distance due to the frequency dependent faster decay of the $2\omega$ amplitude. Even further outside the hotspot region, the amplitude ratio seems to decrease again, which can clearly be attributed to the influence of the white phase noise when the signal gets too small.

\begin{figure*}
    \centering
    \includegraphics[width =\textwidth]{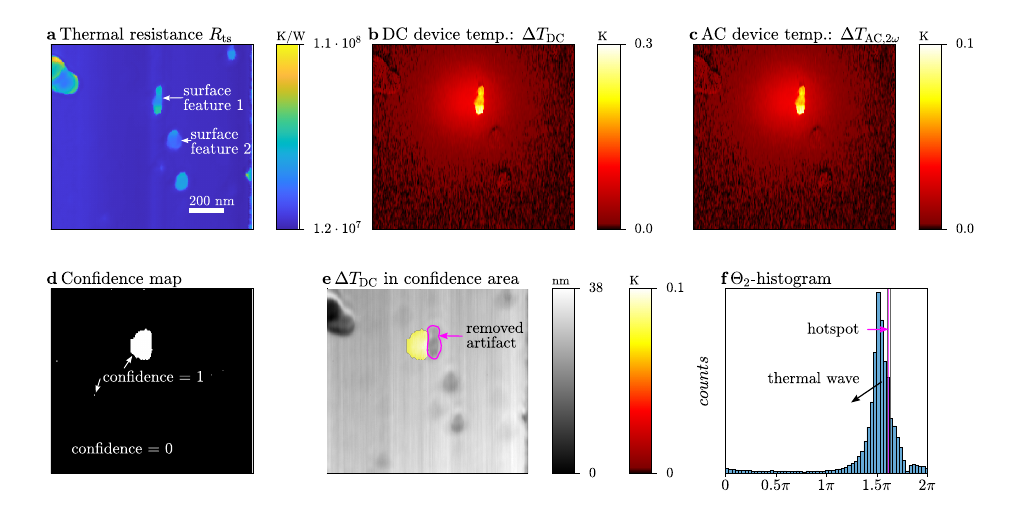}
    \caption{Tip-sample thermal resistance $\Rts$ (a), DC and AC temperature rise above room temperature in K (b,c). Confidence map, where 0 is equivalent to not trustworthy, and 1 is equivalent to trustworthy (d). DC temperature rise in the confidence area, in overlay with topography (e). Phase-histogram of $\Theta_2$ indicating the trusted interval of phases and the decay due to thermal waves}
    \label{fig:Nonlineardevice_Tanalysis_confidence_map}
\end{figure*}

In the topography images, several protrusions are visible which consist of residues from clean-room fabrication. Two of these lie in the region of the hotspot. At the protrusions, the amplitudes $\dVacOne$ and $\dVacTwo$ are low and the phase signals are delayed which is consistent with contamination by particles of low thermal conductivity or poor thermal coupling to the electrode surface.
The calculated thermal resistance signal $\Rts$ and temperature rises $\dTdc$ and $\dTacTwo$ are shown in figure \ref{fig:Nonlineardevice_Tanalysis_confidence_map} a-c. The calculated $\Rts$ spans one order of magnitude from $1.2\cdot 10^7\,\mathrm{K/W}$ to $1.1\cdot 10^8\,\mathrm{K/W}$, which illustrates the general challenge of nanoscale thermometry: the large contrast in thermal resistance.

The second particle, which is indicated in figure \ref{fig:Nonlineardevice_Tanalysis_confidence_map} a, shows in an exemplary way that the method described here can correct for strong variations in thermal resistance. Despite the low thermal conductance of the particle, the particle is hardly visible in the extracted temperature map, in agreement with the expectation that the passive particle adopts the temperature of the underlying device region. This also indicates clearly that a change of sample properties through interaction with a hot probing tip is not significant. As an experimental indication, the clean phase signal in the particle area (figure \ref{fig:Nonlinear_device_SThM_raw_data} e and f) is a good measure, only showing deviations at the edges, where the signal changes too suddenly for the measurement bandwidth.    

Turning to the first particle, we present an example where the analysis assumptions do no longer hold. 
The tip-sample thermal resistance is particularly large in this area. In this case, the calculated temperature rises $\dTdc$ and $\dTacTwo$ are much larger than those in the underlying hotspot region. 
In the $\Theta_2$ phase-signal, this artifact can be clearly identified: it has a distinctly different phase from the center region of the hotspot, and does not show the same decline of the phase as the surrounding of the hotspot does. 

Consequently, we can use the phase information to create a confidence map and separate the erroneous region. To do so, the phase of the center of the hotspot of $1.61\pi$ is chosen as the trusted phase, and a phase deviation of $0.01\pi$ is chosen as a threshold, as shown in figure \ref{fig:Nonlineardevice_Tanalysis_confidence_map} f. Then, a 3x3 filter is applied to each pixel, where each pixel is decided to be trustworthy if at least 5 of the pixels in the grid are within the trusted phase interval, creating a confidence map as depicted in subfigure d. This additionally robustly removes pixels in which the amplitude signal is approaching the noise level, such that the temperature can not be calculated correctly. The resulting trusted region of the DC temperature rise is shown in subfigure e, where the removed artifact is indicated by a pink circle. The remaining trusted area consists of the hotspot and few, small areas throughout the scan with a locked phase. Since the choice of narrow phase-filtering has cut out the radial heat spreading in the surrounding of the hotspot, the data depicted in the main text in figure \ref{fig:Nonlinear_device_SThM_images} d and e contains all vertical line profiles to the left of the hotspot center instead of the confidence area. Regardless, we stress the possibility of using the phase signal for detecting not-trustworthy areas, resulting from both artifacts and regions of signal close to the noise floor, and filtering them out.

\subsection{Estimation of filament width}

Due to the filament being buried beneath the thermally well conducting top electrode, the properties of the filament are not directly accessible. However, making use of a simplified, analytical model of the heat-flow within RRAM cells and comparing the result with the measured temperature distribution, some boundaries to the parameters of the system can be identified.

According to Yovanovich and Marotta\cite{Yovanovich2006}, the temperature distribution at a given height $z$ and radial distance $r$ from an isothermal, circular heat source in a semi-infinite half-space has an analytical solution and can be described as follows 
\begin{align}
    T_{\mathrm{TE}}(r,z) &= \frac{2}{\pi} T_{\mathrm{TE},1} \nonumber \\ &  \cdot \sin^{-1} \left( \frac{2\cdot a}{\sqrt{(r-a)^2 +z^2} + \sqrt{(r+a)^2 +z^2} } \right) \label{eq:T_of_r_z}
\end{align}
The calculated distribution depends only on geometrical factors, but not on material parameters: the radius of the heat-source, $a$, and the temperature of the heat-source, $T_{\mathrm{TE},1}$. 
Assuming that the temperature distribution on top of the finite electrode can be described well by the temperature distribution according to \eqref{eq:T_of_r_z} at the height of the top electrode thickness, one can use the analytical solution to extract the radius of the filament. That this approximation gives a reasonable indication is easily verified using finite element modeling.

\begin{figure}[hbt!]
\includegraphics[width=0.5\textwidth]{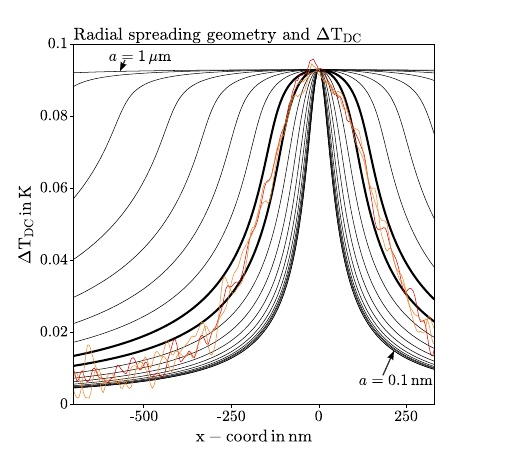}
\caption{Temperature profiles in radial spreading geometry with isothermal circular heat source of radius $a$ according to \eqref{eq:T_of_r_z} with radii of $0.1\,\mathrm{nm}$ and $1\,\mathrm{nm}$, then in finer logarithmic spacing from $10\,\mathrm{nm}$ to $1\,\mathrm{\mu m}$ in black, thickened for $100\,\mathrm{nm}$ and $133\,\mathrm{nm}$. Measured temperature profile through the center of the hotspot in red and the neighbouring 2 profiles in orange.}
\label{fig:RRAM_T_profiles_radius}
\end{figure}

The experimental temperature data is shown in figure \ref{fig:RRAM_T_profiles_radius}. The  central line profile along the fast scan axis through the hotspot is indicated in red. The orange lines indicate the two neighbouring temperature profiles. The black solid lines indicate the expected temperature distributions for assumed filament diameters from $0.1\,\mathrm{nm}-1\,\mathrm{nm}$ and in logarithmic spacing from $10\,\mathrm{nm}-1\,\mathrm{\mu m}$. The two thick black lines give the best fit to the overall shape of the central profile, which yields a heat-source radius between $100\,\mathrm{nm}$ and $133\,\mathrm{nm}$. We note, however, that around the center the experimental temperature profile is less flat than predicted. This could be indicative of a limitation of such modelling describing the filament itself in a simplified manner.
However, even when taking into account other filament models, the heat-source radius extracted here is either extracted correctly or overestimated, such that the extracted value gives an upper boundary of the heat-source radius in the studied RRAM cell.

\bibliography{bibliography.bib}
\end{document}